%% file: proceedings.tex
\documentclass{sigchi}
\usepackage{balance}  
\usepackage{graphicx}
\usepackage{txfonts}
\usepackage{times}  
\usepackage[pdftex]{hyperref} 
\usepackage{color}
\usepackage{textcomp}
\usepackage{booktabs}
\usepackage{ccicons}
\usepackage{todonotes}
\usepackage{xspace}
\usepackage{enumitem}
\usepackage{caption}
\usepackage{subcaption}

\newcommand{\cf}{\fontfamily{cmss}\selectfont\xspace}
\makeatletter
\def\url@leostyle{%
  \@ifundefined{selectfont}{\def\UrlFont{\sf}}{\def\UrlFont{\small\bf\ttfamily}}}
\AtBeginDocument{  
    \newcommand*{\sref}[1]{\textit{
                \S\nameref{#1}
                }}
}
\makeatother
\def\pprw{8.5in}
\def\pprh{11in}

\definecolor{tangerine}{rgb}{1.0, 0.6, 0}
\definecolor{ocean}{rgb}{0.2, 0.4, 0.5}

\definecolor{light-gray}{gray}{0.5}
\definecolor{linkColor}{RGB}{6,125,233}
\hypersetup{%
  pdftitle={Redistributing Funds across Crowdfunding Campaigns},
  pdfauthor={Abouzied et al.},
  pdfkeywords={Donation Packages; Giving Clubs; Fund Redistribution; Crowdfunding Platforms},
  bookmarksnumbered,
  pdfstartview={FitH},
  colorlinks,
  citecolor=black,
  filecolor=black,
  linkcolor=black,
  urlcolor=linkColor,
  breaklinks=true,
}

\newcommand{\lag}{{LaunchGood}\xspace}

\newcommand{\cpr}{{choice-preserving redistribution}\xspace}
\newcommand{\car}{{choice-agnostic redistribution}\xspace}
\newcommand{\Cpr}{{Choice-preserving redistribution}\xspace}
\newcommand{\Car}{{Choice-agnostic redistribution}\xspace}

\usepackage{array}
\newcolumntype{P}[1]{>{\centering\arraybackslash}p{#1}}

\begin{document}


\title{Redistributing Funds across Charitable Crowdfunding Campaigns}

\numberofauthors{3}
\author{%
 \alignauthor{Matteo Brucato\\
    \affaddr{UMass, Amherst}\\
    \email{matteo@cs.umass.edu}}\\
 \alignauthor{Azza Abouzied}\\
    \affaddr{NYU, Abu Dhabi}\\
    \email{azza@nyu.edu}\\
 \alignauthor{Chris Blauvelt\\
    \affaddr{LaunchGood}\\
    \email{chris@launchgood.com}} 
}

\maketitle

\begin{abstract}

On Kickstarter only 36\% of crowdfunding campaigns successfully raise sufficient funds for their projects. In this paper, we explore the possibility of redistribution of crowdfunding donations to increase the chances of success. We define several intuitive redistribution policies and, using data from a real crowdfunding platform, \lag, we assess the potential improvement in campaign fundraising success rates. We find that an aggressive redistribution scheme can boost campaign success rates from 37\% to 79\%, but such \textit{\car schemes} come at the cost of disregarding donor preferences. Taking inspiration from offline giving societies and donor clubs, we build a case for \textit{\cpr schemes} that strike a balance between increasing the number of successful campaigns and respecting giving preference. We find that \cpr can easily achieve campaign success rates of 48\%. Finally, we discuss the implications of these different redistribution schemes for the various stakeholders in the crowdfunding ecosystem.

\end{abstract}

\keywords{Fund Redistribution; Crowdfunding Platforms; Giving Clubs}

\category{H.5.3.}{Information Interfaces and Presentation}{Group and Organization Interfaces}

\input{intro.tex}
\input{relatedwork.tex}

\input{case.tex}
\input{discussion.tex}

\input{conclusion.tex}

\section{Acknowledgments}
This material is based upon work supported by the National Science Foundation under grant IIS-1420941.

\balance{}

\bibliographystyle{SIGCHI-Reference-Format}
\bibliography{refs}

\end{document}

%% file: intro.tex
\section{Introduction}

Crowdfunding provides individuals and organizations with the opportunity to raise funds for 
innovative projects, charitable causes, or public services. By raising the visibility of these fund-raising campaigns to Internet scale, crowdfunding sites like Kickstarter, Indiegogo, and Donorschoose dramatically increase the pool of potential backers and chances of successful fund-raising. Despite this potential, most crowdfunding campaigns still fail to reach their funding goals. On Kickstarter, the largest crowdfunding platform, only 36\% of the campaigns, were successfully funded in 2016~\cite{kickstarterstats}.

In 
the prevalent
\textit{all-or-nothing} funding model, campaigns that fail to meet their funding goals receive none of the contributions: even if a campaign attracts 99\% of its funding goal, it will not be funded. Currently, 1445 Kickstarter campaigns are unsuccessful, despite having raised 81-99\% of their goal~\cite{kickstarterstats}. In contrast, campaigns that exceed their goals keep all contributions; `Exploding Kittens' and `Pebble' raised 87,825\% and 4,067\% of their funding goals respectively. This raises the question: \textit{can we redistribute contributions to successfully fund more campaigns?}

Clearly, a redistribution where excess funds from over-funded campaigns are given to under-funded ones, can lead to an overall increase in successful campaigns. A na\"ive redistribution, however, can be detrimental to both the overall quality of campaigns and to the degree of funding. For example, if campaign organizers believe that extra funds will eventually be redistributed in their favor, they may be less motivated to produce higher quality campaigns; if funders think their donations could end up backing campaigns that they do not like, they may not be willing to contribute.

Therefore, redistribution must be done carefully. Consider, a donor, Sandy, who backs three simultaneous campaigns. Assuming that Sandy wishes all three campaigns to succeed regardless of how other donors give, she may be willing to accept a redistribution of her funds across the three campaigns if it allows them to meet their funding goal. She is less likely to accept a redistribution that allocates her funds to other campaigns that she does not support. Redistribution schemes may either honor Sandy's giving preferences, by only redistributing her funds among the campaigns she has contributed to (\textit{\cpr schemes}) or ignore Sandy's preferences and redistribute funds to any active campaign (\textit{\car schemes}).

In this paper we explore several possible crowdfunding donation redistribution schemes. We begin by defining several archetypes that represent intuitive redistribution policies (\sref{sec:model}). We then analyze the potential improvements to \textit{efficiency} --- measured in terms of campaigns successfully funded\footnote{Viewing crowdfunding platforms as online markets where the goal is to match donor contributions to successful campaigns, the percentage of successful campaigns within a platform is a good approximation of its efficiency~\cite{wash-returnrule}.} --- achievable by these redistribution schemes using real crowdfunding data from \lag. We find that an aggressive \car can boost campaign success rates from 37\% to 79\%, but such a policy comes at the cost of completely ignoring donor preferences. We build a case for \cpr policies that strike a balance between increasing successful campaigns from 37\% to 48\% and respecting giving preferences (\sref{sec:results}). Finally, we discuss the potential implications of various redistribution policies when implemented on crowdfunding platforms (\sref{sec:discussion}). We contextualize our work within the existing research on crowdfunding platforms that aim to improve success outcomes for campaigns and platforms (\sref{sec:background}). 

%% file: relatedwork.tex
\section{Background}
\label{sec:background}

The goal of all crowdfunding platforms is generally the same: to match a campaign's needs with funding from donors. Campaigns are only successful if a sufficient number of donors \textit{coordinate} to contribute to a project. 

\subsection{Crowdfunding Mechanisms}

Crowdfunding campaigns are often used to fund discrete public goods that require a certain amount of money to be raised to be useful~\cite{belleflamme, greenberg2012crowdfunding, wash-returnrule}. Kickstarter employs an \textit{all-or-nothing} model or \textit{return rule} where a project collects money only if the funding goal is met. The contrasting model supported by IndieGoGo is a \textit{keep-it-all} or \textit{direct donation} model where donations are retained even if the funding goal is not met. These crowdfunding mechanisms have been shown to have a significant impact on donor perceptions of campaigns, donor willingness to contribute, and the eventual success of campaigns~\cite{cumming2014crowdfunding, wash-returnrule}.

Specifically, Wash and Solomon showed through a series of controlled experiments that the return rule increases donors' willingness to donate to riskier projects and thus more accurately reflects individual preferences rather than the funding of projects that are more likely to be funded~\cite{wash-returnrule}. However, while more projects are successfully funded through the return rule, this benefit comes with a reduction in efficiency 
when individuals donate to their own preferences rather than those of the crowd.

\textit{These results motivate our work on understanding whether mechanisms like redistribution can simultaneously increase the number of successfully funded projects while respecting individual preferences.}

\subsection{Social Proof and Coordination}

Crowdfunding campaigns rely heavily on social proof. Displaying information like total funds raised and the number of donors can signal to the individual the collective valuation of the crowd. Participation by other donors signals the quality and credibility of the campaign, removes apprehensions, provides evidence of reciprocity, and establishes norms for how much to donate. These signals provide critical evidence that helps donors coordinate around supporting high value campaigns~\cite{mollick2014dynamics, cotterill2011impacts}.

Unfortunately, crowdfunding platforms do not always give donors accurate information about the crowd's beliefs~\cite{wash-time}: the crowd's valuation of a project induced by donation amounts on crowdfunding sites may be (i) delayed, 
(ii) misrepresented, 
or (iii) overshadowed by other projects~\cite{wash-time, andreoni1988free, wash-skew}.\footnote{Solomon et al. show that highly successful star projects can actually hinder other projects~\cite{wash-skew}.} As a result, campaigns can fail to reach their funding goal and the overall distribution of donations may not resemble the crowd's actual valuations. Codo~\cite{codo} is one crowdfunding system that can potentially mitigate this issue by allowing individual donors to make independent valuations that are stipulated on the crowd's valuations. 

In relation to these works, \textit{donation redistribution can be construed as a coordination of donors' funds}.

\subsection{Improving Outcomes}

Recent research investigated several different factors influencing crowdfunding outcomes. Gerber et al. describe the motivations of both campaign organizers and donors for using crowdfunding as a fundraising tool~\cite{Gerber:2013:CMD:2562181.2530540}. Hui et al. explore the role that a crowdfunding project's community can play in its success~\cite{Hui:2014:URC:2531602.2531715}. It has also been demonstrated that the use of persuasive language~\cite{Mitra:2014:LGP:2531602.2531656} and status updates~\cite{xu2014show} can influence the chances of crowdfunding success. 

Social networks can also be leveraged to increase donations. A donation request or tagging from family or others who have donated can increase individual contributions~\cite{cotterill2011impacts}. Social networks can be used by organizers to improve fundraising outcomes if the existing networks can be activated and expanded~\cite{Hui:2014:ULS:2598510.2598539}. 
Finally, the timing of donations is a factor that affects donations~\cite{wash-time}, and donors are more willing to donate more to projects that are nearing completion~\cite{complete-effect}.

Real crowdfunding sites use a variety of incentives for donors to donate. Direct inducements include gifts, stretch goals, early access, and even equity.

\textit{Our work explores several possible automated reallocation schemes that can improve outcomes for both campaign organizers and donors.}

\subsection{Redistribution in the Wild}

Giving societies or donor clubs are organizations that explicitly gather funds from donors and allocates them toward specific projects. These giving societies are usually unified under high-level causes (e.g., education, environmental conservation, etc.). In this manner, the donation process is re-framed from being a highly personal choice to a collective action. The donor slightly relaxes their personal autonomy to defer to the valuations and expertise of the collective. Giving clubs fit within our \cpr schemes as donors implicitly agree to a redistribution of their funds across the subset of campaigns managed by the giving club.

 Donorschoose is a crowdfunding site focused on education that specifically targets high-poverty public schools to increase classroom engagement. Success rates are relatively high on DonorsChoose ($\approx 74\%$). Projects are vetted by the website staff and funds to an unsuccessful project are not refunded, but repurposed. Donors can choose alternate projects to fund instead or allow the platform operators or the teacher whose campaign was unsuccessful to repurpose the funds as she sees fit. \textit{We compare repurposing to other redistribution effects.}

%% file: case.tex
\section{Empirical Case for Fund Redistribution}

In this section, we describe how we implement and analyze two classes of redistribution schemes: \car and \cpr. We study the efficiency models with real data from the \lag crowdfunding platform.

\subsection{The \lag Dataset}
\label{sec:data}

\textit{\lag} is a niche crowdfunding platform focused on the Muslim community. \lag went live in October 2013. Since its inception, \lag has raised more than 3 million dollars for more than 300 projects. Among its notable campaigns are a campaign to rebuild African-American churches destroyed by arson in 2015 and a campaign to fund Adnan Syed's legal team, both raising more than 100,000 dollars.

We analyze \lag's entire donation trace since its inception. We eliminate active campaigns from our analysis (i.e. campaigns whose end date is after our snapshot date of January, 2016) and campaigns that remained live for more than seven days after their end date\footnote{It is difficult to determine whether campaigns overlap, if they have no set end date.}. \lag keeps track of offline contributions made by anonymous donors directly to the campaign organizers. We cannot redistribute these funds as they are not raised through the platform.

\begin{figure}[htbp]
    \centering
    \includegraphics[width=1\linewidth]{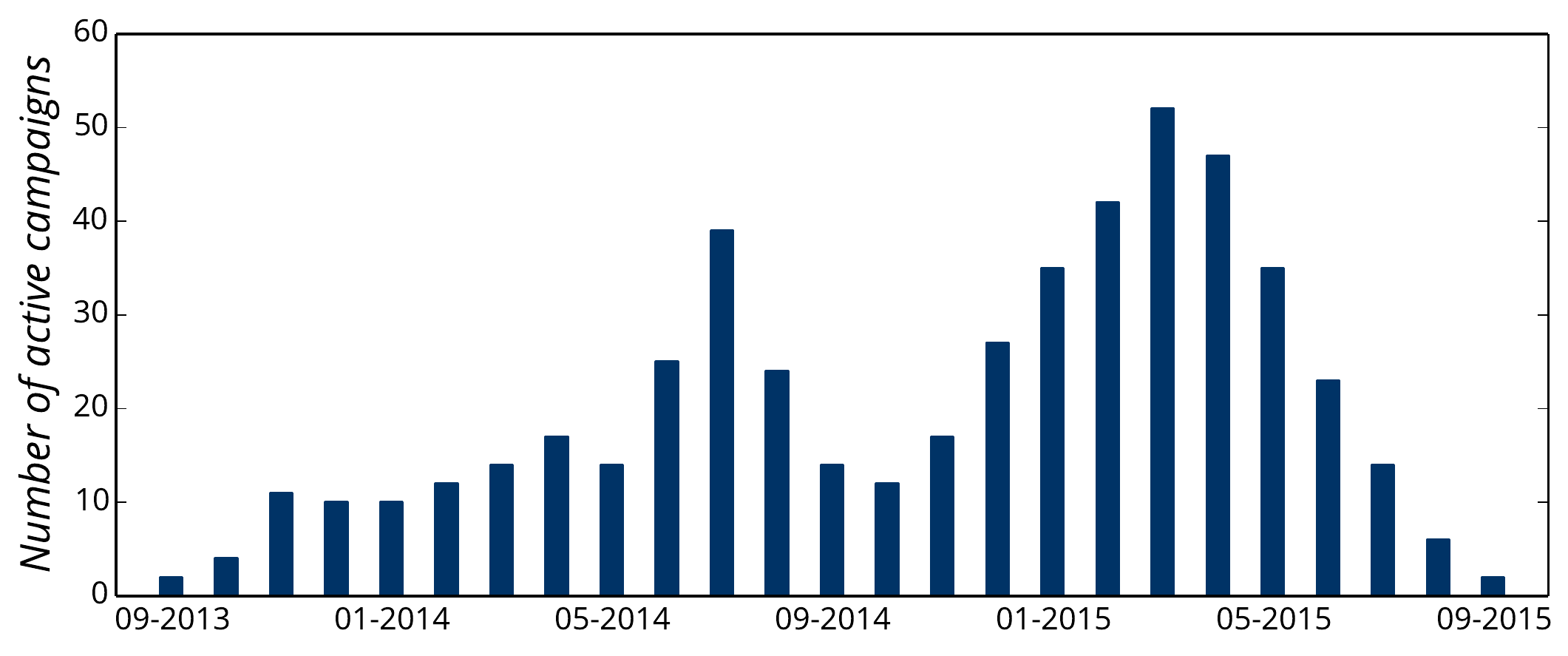}
    \caption{Number of active campaigns per month.}
    \label{fig:over-time}
\end{figure}

\begin{figure}[htbp]
    \centering
    \includegraphics[width=1\linewidth]{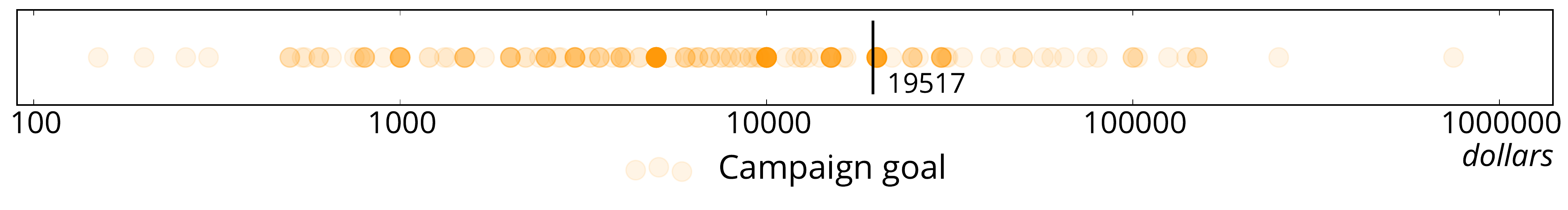}
    \caption{A dot-plot of campaign goals. The mean campaign goal marked with the vertical black line is 19,517 USD.}
    \label{fig:goals}
\end{figure}

Figure \ref{fig:over-time} illustrates the number of active campaigns per month. Figure \ref{fig:goals} charts the distribution of campaign goals.

\begin{table}[htpb]
    {\cf \small
    \centering
    \begin{tabular}{p{0.79\linewidth}c}
        \toprule
        \textbf{Property} &  \\
        \midrule
        Total number of campaigns & 228 \\
        Distinct donors & 7935 \\
        Repeat donors & 1342 \\
        Percentage of repeat donors &  16.9\% \\
        Average campaigns backed per repeat donor & 3.7  \\
        Maximum campaigns backed per repeat donor & 142 \\
        Average time interval between consecutive donations & 96 days \\
        Average life span of a campaign & 44 days \\
         \bottomrule
    \end{tabular}
    }
    \caption{Summary Statistics for \lag's Campaigns \& Donors.}
    \label{tab:donorstats}
\end{table}

\begin{figure*}
    \centering
    \includegraphics[width=1\linewidth]{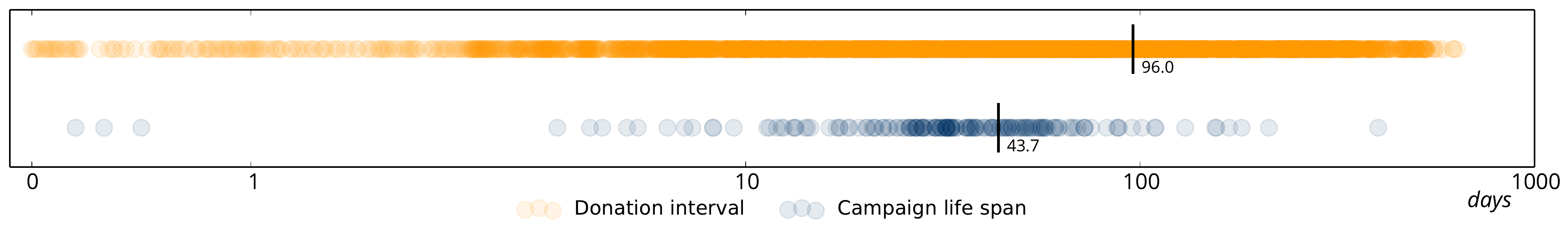}
    \caption{Each {\color{tangerine} tangerine} dot represents the time period between two consecutive donations by the same donor to different campaigns. Each {\color{ocean} ocean} dot represents the life span of a campaign.}
    \label{fig:span}
\end{figure*}

\input{mlp.tex}

\subsection{Results}
\label{sec:results}

Table \ref{tab:overall} shows the increase in campaigns funded after each redistribution scheme. Except for a single campaign that was successfully funded by \cpr and failed by \car, the set of campaigns funded by ordered \cpr is a subset of unordered \cpr, which is a subset of repurposing, which in turn is a subset of na\"ive redistribution.

\begin{table}[htpb]
    {\cf \small
    \centering
    \begin{tabular}{p{0.74\linewidth}cc}
        \toprule
        \textbf{Projects funded with ... } &  \\
        \midrule
        Original contributions (no redistribution) & 85 & 37\% \\
        \midrule
        \textit{\Car} & \\
        \ \ \ \ Na\"ive & 180 & 79\% \\  
        \ \ \ \ Repurposing & 175 & 77\% \\
        \midrule
        \textit{\Cpr} & \\
        \ \ \ \ Unordered & 109 & 48\% \\
        \ \ \ \ Ordered & 99 & 43\% \\
        \bottomrule
    \end{tabular}
    }
    \caption{Successful campaigns for each redistribution scheme.}
    \label{tab:overall}
\end{table}

The increase in successful campaigns with \cpr schemes is low when compared to \car schemes. Yet, if we consider (i) the small proportion of repeat donors, roughly 17\% of the donor base (Table \ref{tab:donorstats}), (ii) the mean number of campaigns a repeat donor contributes to, less than 4 campaigns (Table \ref{tab:donorstats}), and (iii) the mean gap between consecutive donations, 96 days, in relation to the average life span of campaigns, 44 days (Figure \ref{fig:span}), the increases in campaign success brought about by \cpr are actually fairly surprising.

\subsubsection{Trade off: Efficiency vs. Choice}

\begin{figure*}[htbp]
    \centering
    \includegraphics[width=1\linewidth]{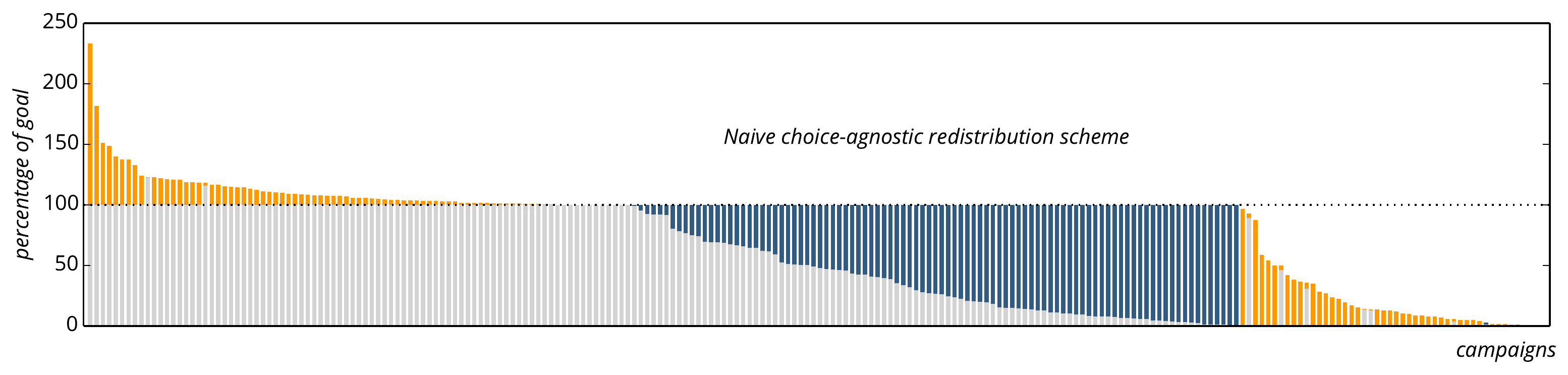}
    \includegraphics[width=1\linewidth]{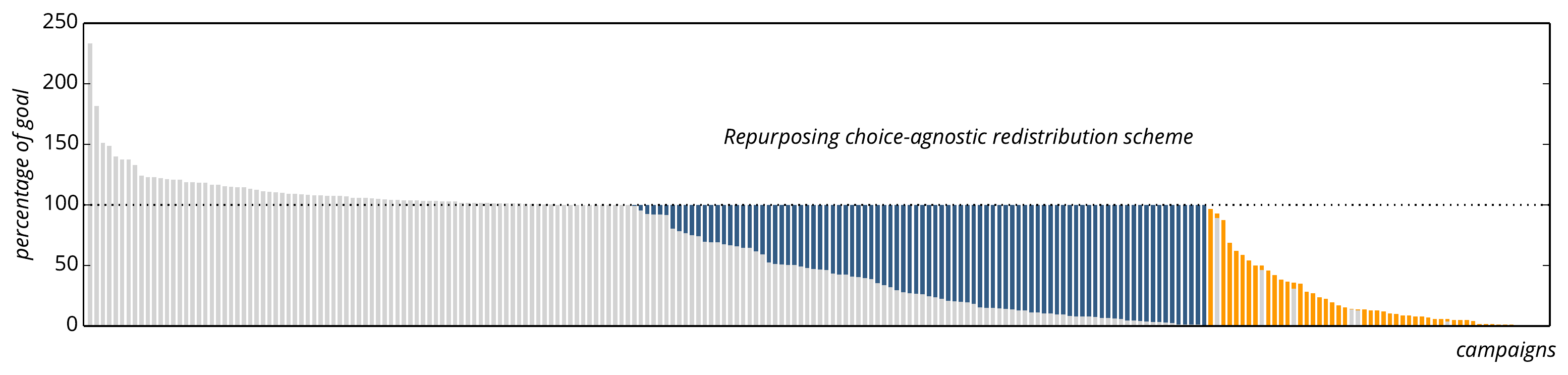}
    \caption{The behaviour of \car schemes with \lag's contributions. The {\color{tangerine} tangerine} colored bars indicate funds deducted 
    from a campaign. The {\color{ocean} ocean} colored bars indicate funds allocated to a campaign after redistribution.}
    \label{fig:inter}
\end{figure*}

Naturally, \car leads to more efficiency in terms of campaigns meeting their goal 
than \cpr.

\begin{figure*}[htbp]
    \centering
    \includegraphics[width=1\linewidth]{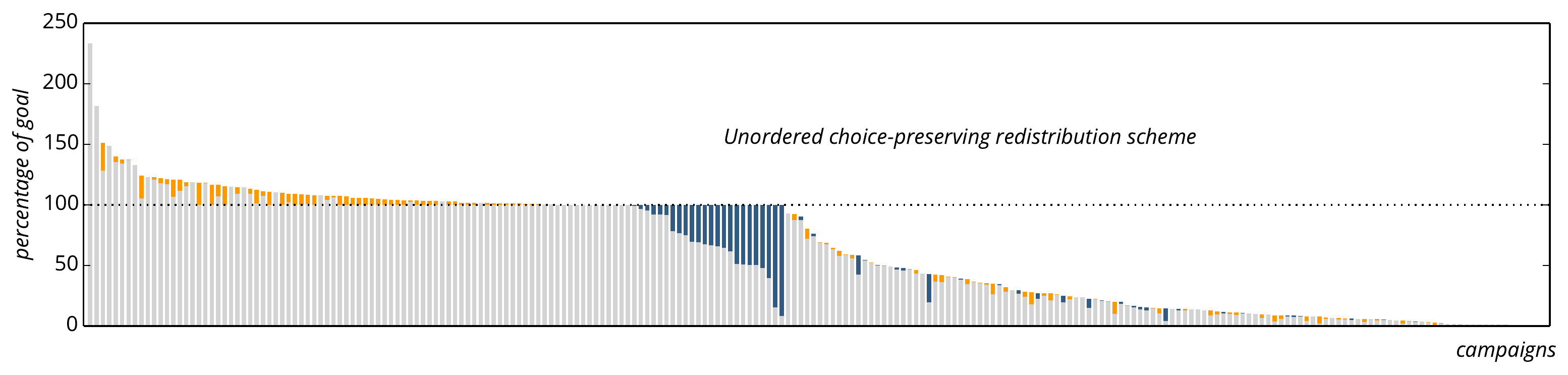}
    \includegraphics[width=1\linewidth]{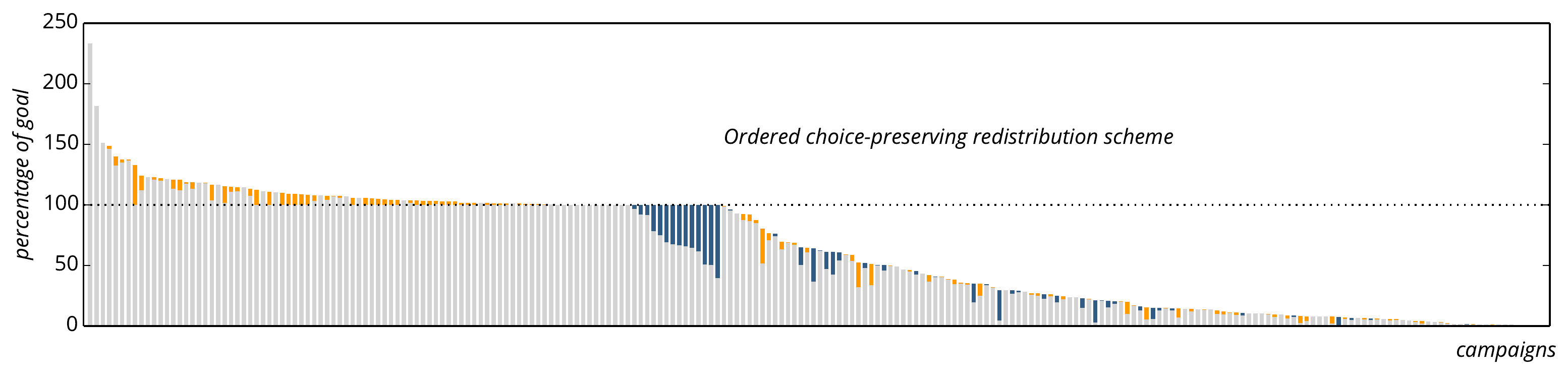}
    \caption{The behaviour of \cpr schemes with \lag's contributions. The {\color{tangerine} tangerine} colored bars indicate funds deducted 
    from a campaign. The {\color{ocean} ocean} colored bars indicate funds allocated to a campaign after redistribution.}
    \label{fig:intra}
\end{figure*}

Figure \ref{fig:inter} illustrates the behavior of \car schemes. With na\"ive redistribution, almost all over-funded campaigns lose most of their excess funds. We observe that repurposing is slightly less efficient than na\"ive redistribution as it allows successful campaigns to keep their excess funds. This contrasts with the behavior of \cpr schemes (Figure \ref{fig:intra}), where only a few over-funded campaigns lose some of their excess funds, but donor choices (Fig. \ref{fig:intra}-top) or ordered preferences (Fig. \ref{fig:intra}-bottom) are preserved.

In general, our definition of \car favors campaigns with small goals and disfavors campaigns with large goals regardless of how much of their goals they initially raised. The short tail section, the noticeable number of post-redistribution successful campaigns that initially raised less than 10\% of their goal and the noticeable number of failed campaigns that initially raised more than 50\% of their goal in Figure \ref{fig:inter} illustrate this behavior. Table \ref{tab:diff} also shows that on average \car supports campaigns that are further from their goal.\footnote{As a redistribution scheme funds more campaigns the average difference from the goal naturally increases and becomes closer to the distribution of differences of failed campaigns with original contributions. Therefore, this measure should be interpreted carefully.}

\begin{table}[htpb]
    {\cf \small
    \centering
    \begin{tabular}{p{0.5\linewidth}P{0.125\linewidth}P{0.2\linewidth}}
        \toprule
        \textbf{Across ... } & \multicolumn{2}{c}{\textbf{Average}} \\
        & goal & diff. from goal \\
        \midrule

All campaigns 	&	19517 & 11548 (40\%) \\
Only successful campaigns &	11221 & -1090 (-12\%)  \\

Only failed campaigns & 24448 & 19059 (70\%)\\
\midrule
\multicolumn{3}{l}{\textit{Only successful campaigns after ... redistribution}} \\

        Na\"ive 	 &	5749 & 	3580 (66\%) \\
        Repurposing &	5337 &	3312 (66\%) \\
        Unordered choice-preserving &	6429&		1357 (34\%) \\
Ordered choice-preserving &	6411 &	1455 (31\%)\\
        \bottomrule
    \end{tabular}
    }
    \caption{Summary statistics on differences from goal across successful campaigns under different redistribution schemes.}
    \label{tab:diff}
\end{table}

This behavior is most clearly manifested in one redistribution outcome where a campaign that raised 97\% of its goal of 26,000 USD, with 800 USD dollars remaining. Both na\"ive redistribution and repurposing fail this campaign to support others. Both \cpr schemes, however, help this campaign meet its goal.

This may seem like a shortcoming of the optimization objective, which staunchly maximizes the number of campaigns funded. This is an artifact of our optimization objective. If different resolution properties are desired, platform designers could change the optimization problem to also minimize the sum of (absolute/percentage) differences reallocated. Designers must also weigh maximizing success against minimizing differences, where each weight assignment leads to different outcomes.

\Cpr is biased toward helping campaigns with multiple donors as they are more likely to benefit from a redistribution of funds from their supporting donor base. The average percentage of repeat donors (i.e. donors who also donated to other campaigns) per campaign for the successful campaigns post \cpr is 61-63\%. This is much higher than the overall average percentage of repeat donors per campaign 48\% (Table \ref{tab:repeat}). The one campaign that only \cpr schemes successfully funded had 120 donors, 64 (53\%) of them were repeat donors.

\begin{table}[htpb]
    {\cf \small
    \centering
    \begin{tabular}{p{0.5\linewidth}P{0.125\linewidth}P{0.2\linewidth}}
        \toprule
        
        \textbf{Across ... } & \multicolumn{2}{c}{\textbf{Average}} \\
        
        & num. of donors & num. of repeat donors \\
        \midrule
        All campaigns	& 51 & 22 (48\%)	 \\
        Only successful campaigns & 	87 	& 35 (47\%)	\\
        Only failed campaigns	& 29 & 14 (49\%)	\\
        \midrule
        \multicolumn{3}{l}{\textit{Only successful campaigns after ... redistribution}} \\
        Na\"ive & 22 & 11 (50\%)\\
        Repurposing &	21 & 11 (50\%) \\
        Unordered choice-preserving &	47 & 29 (61\%) \\
        Ordered choice-preserving & 55 & 38 (63\%) \\      
        \bottomrule
    \end{tabular}
    }
    \caption{Summary repeat donor statistics across successful campaigns under different redistribution schemes.}
    \label{tab:repeat}
\end{table}

\subsubsection{Effect of Donor Acceptance}
\label{sec:acceptance}

\begin{figure*}[htbp]
    \centering
    \begin{subfigure}[t]{1\columnwidth}
        \includegraphics[width=1\columnwidth]{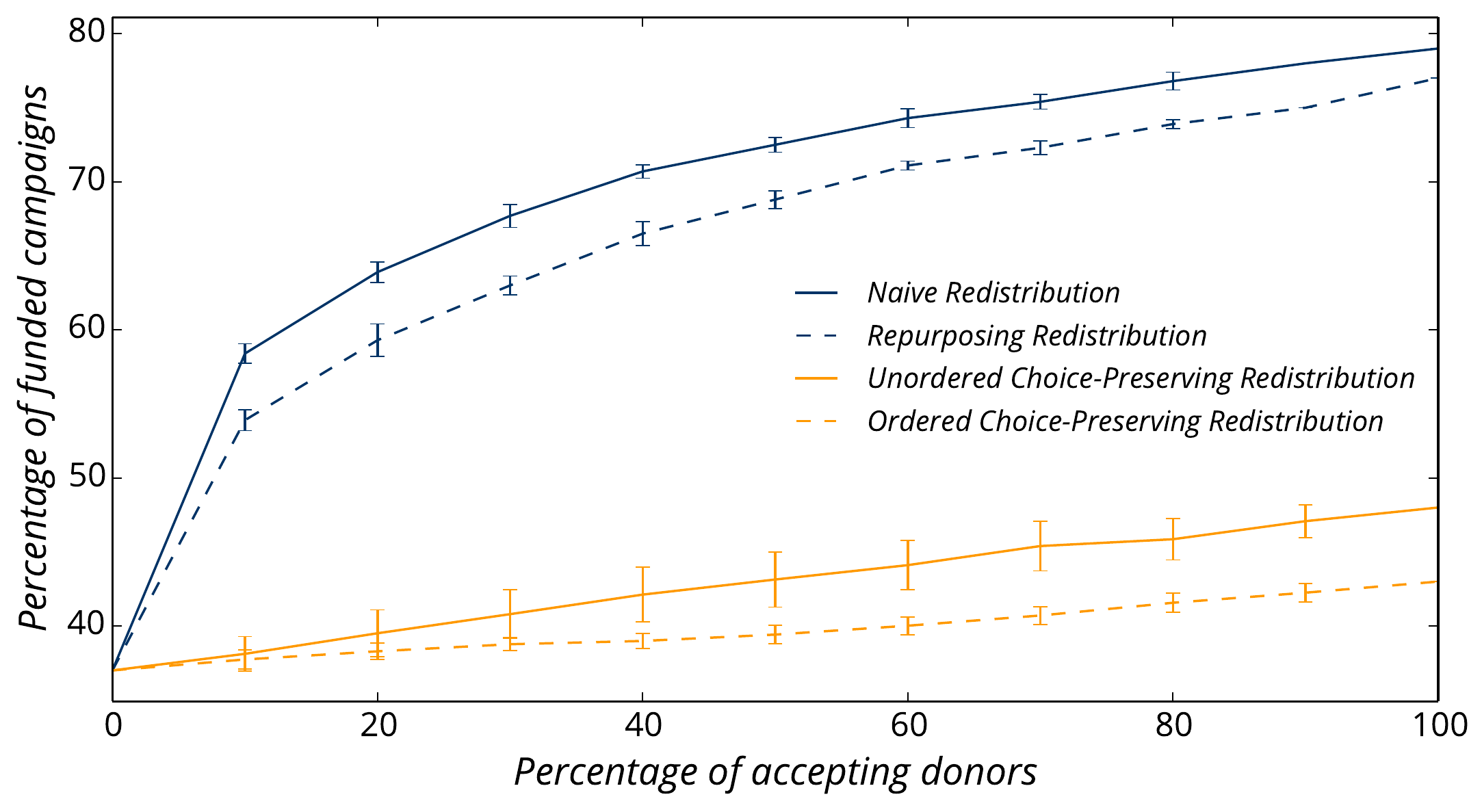}
        \caption{The mean increase in funded campaigns as the percentage of donors accept a redistribution scheme. 
        }
        \label{fig:accept}
    \end{subfigure}
    \hspace*{\fill}%
    \begin{subfigure}[t]{1\columnwidth}
        \includegraphics[width=1\columnwidth]{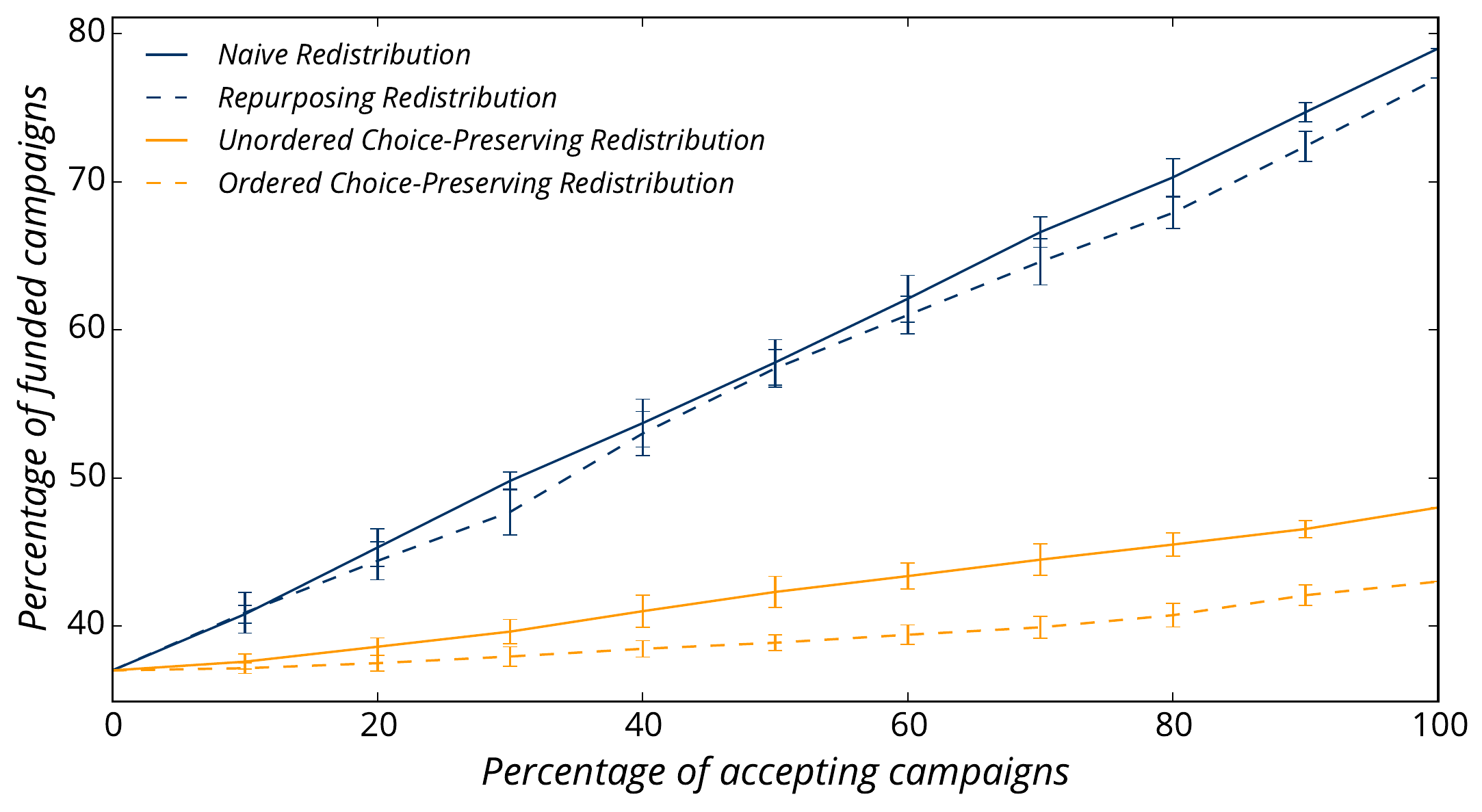}
        \caption{The mean increase in funded campaigns as the percentage of campaigns accept a redistribution scheme.
        }
        \label{fig:org-accept}
    \end{subfigure}
    \caption{The mean increase in funded campaigns as the percentage of donors (a) or campaigns (b) accept a redistribution scheme.
    For each acceptance percentage, we construct random samples of donors (a) or campaigns (b) who accept the redistribution (both in terms of outflow or inflow of funds) 
    to estimate the variance: the error bars mark one standard deviation from the plotted mean.}
\end{figure*}

We simulated the behavior of each scheme under different donor acceptance ratios as follows: for each acceptance percentage $p\%$, we randomly picked $p\%$ of the donor population to agree to a redistribution of their funds. The remaining $(100 - p)\%$ rejects any redistribution. We generated several samples for each $p\%$ to compute the mean number of successful campaigns and the variance for a given $p\%$. 

Figure \ref{fig:accept} illustrates the effect increasing the number of donors who accept a redistribution on efficiency. 

With \car, an acceptance ratio of only 10\% of the donors can lead to almost half of the improvement achievable when all donors accept the redistribution. In contrast, \cpr leads to a more gradual improvement as the acceptance ratio increases.

\subsubsection{Effect of Organizer Acceptance}
As with donor acceptance rates, we also simulated a percentage of campaigns organizers accepting a redistribution of funds to their campaigns, both in terms of outflow and inflow of funds.

Figure \ref{fig:org-accept} illustrates the effect increasing the number of organizers who accept a redistribution on efficiency. In this case, both \car and \cpr show the same linear behavior in response to increases of accepting organizers.

%% file: mlp.tex
\subsection{Redistribution Schemes}
\label{sec:model}

We analyze four redistribution schemes broadly classified into \textit{\car} or \textit{\cpr} schemes. \Car schemes allow a donor's contributions to be redistributed to any campaign within the platform. These include \textit{na\"ive redistribution} and \textit{repurposing} schemes. \Cpr schemes only shuffle a donor's contributions within the set of campaigns the donor contributed to. These include both \textit{unordered-} and \textit{ordered-} \cpr schemes. With ordering, \cpr ensures that if a donor contributes more to one campaign over another then even after a redistribution, he/she would still contribute more to that campaign.

We formally define each of these schemes within the framework of an optimization problem where the goal is to maximize the number of campaigns that meet their goals.


Let $n$ be the number of distinct donors that donated on the \lag platform and $m$ be the number of distinct campaigns. 

We represent an actual contribution a donor $i \in {1,..., n}$ made to a campaign $j \in {1,..., m}$ with $A_{i,j}$. A campaign, $j$, has a fund-raising goal of $G_j$. We represent whether a campaign has met its goal with a success indicator variable, $I_j$.
\[I_j =
\begin{cases}
    1, & \text{if} \ \displaystyle \sum_{i=1}^{n} A_{i, j} \geq G_j\\
    0, & \text{otherwise}
\end{cases}\]

Every campaign has a start date $s_j$ and an end date $e_j$. Each contribution $A_{i,j}$ is made on a specific date\footnote{If a donor, $i$, makes multiple contributions to the same campaign, $j$, we encode only one contribution $A_{i,j}$ equal to the sum of all such contributions and set $d_{i,j}$ equal to the date of the first contribution.} $d_{i,j}$: $s_j \leq d_{i, j} \leq e_j$.

After a redistribution of funds, we denote the donation a donor $i$ makes to a campaign $j$ with $R_{i, j}$. We now represent whether a campaign meets its goal with another success indicator variable, $I'_j$.
\[I'_j \leq \frac{\sum_{i = 1}^{n} R_{i, j}}{G_j}
\] 

Our redistribution goal is to maximize the number of successful projects,
\[\max \sum_{j = 1}^{m} I'_j \]

\textit{All four schemes must satisfy the following three constraints.} 

\begin{enumerate}[label={[\bf\sc{All\arabic*}]},wide =1em]
\item \emph{Once a winner, always a winner:} If a campaign met its goal without redistribution, a redistribution of funds should not cause this campaign to fail. 
\[\forall j, \;\ I'_j \geq I_j \]

\item \emph{Fixed Budget:} Each donor cannot give more than the sum of his/her original contributions across all campaigns and a donor cannot make a negative contribution.
\[ \forall i, \;\ \sum_{j = 1}^{m} R_{i,j} \leq \sum_{j = 1}^{m} A_{i,j}\]
\[\forall i,j, \;\ R_{i, j} \geq 0\]

\item \emph{Redistribute across live, overlapping campaigns:}
For each contribution $A_{i, j}$ made on $d_{i,j}$, we define overlapping campaigns as follows:
\[\mathcal{O}_{i, j} = \{x : \ e_x \geq d_{i, j} \wedge s_x \leq e_j \}\]





The following constraint ensures that we redistribute funds only within overlapping campaigns.

\[\forall i,j, \;\ \sum_{i, y \in \mathcal{O}_{i, j}} R_{i, y} \leq \sum_{i, x \in \mathcal{O}_{i, j}} A_{i, x}\]


\end{enumerate}

We also try to eliminate the following unfavorable redistributions with the following optional constraints. We drop these \textit{nice} constraints if the optimization problem becomes infeasible:
\begin{enumerate}[label={[\bf\sc{Nice\arabic*}]},wide =1em]
    \item \textit{Avoid allocating more funds to a previously overfunded campaign:}
    \[\forall j, \;\ I_j = 1 \rightarrow \sum_{i=1}^n R_{i, j} \leq \sum_{i=1}^n A_{i, j}\]
    
    \item \textit{Avoid allocating more funds to a failed campaign:} If no redistribution of funds can save a campaign, then we should not allocate additional funds to that campaign. 
    \[\forall j, \;\ I'_j = 0 \rightarrow \sum_{i=1}^n R_{i, j} \leq \sum_{i=1}^n A_{i, j}\]
    
    \item \textit{Avoid surpassing the goal of a previously unsuccessful campaign:} 
    \[\forall j, \;\ I_j = 0 \rightarrow  \sum_{i=1}^n R_{i, j} \leq G_j \]
\end{enumerate}

\subsubsection{\Car}

\paragraph{Na\"ive redistribution} 

This scheme does not require any additional constraints to the base constraints listed above.

\paragraph{Repurposing} 

Repurposing also requires that we only redistribute funds from failed campaigns. We encode this with the following constraint:

\begin{enumerate}[label={[\bf\sc{Rep}]},wide =1em]
    \item \textit{Winners keep all}: 
    \[\forall i,j, \;\ I_j = 1 \rightarrow R_{i, j} = A_{i, j} \]
\end{enumerate}

No other redistribution scheme imposes the ``winners keep all" constraint.

\subsubsection{\Cpr}

Our two \cpr schemes must also ensure the following:

\begin{enumerate}[label={[\bf\sc{CP\arabic*}]},wide =1em]

\item \emph{No new campaigns:} For a given donor, we should only redistribute his/her funds among the campaigns he/she contributed to.
\[\forall i,j, \;\ A_{i, j} = 0 \rightarrow R_{i, j} = 0 \]

\item \emph{Redistribute across live, overlapping campaign contributions:} This constraint tightens the overlap constraints of na\"ive and repurposing redistribution schemes to overlapping \textit{contributions} by the same donor. We redefine overlapping campaign contributions as follows:

\[\mathcal{O}_{i, j} = \{x: \ e_x \geq d_{i, j} \wedge d_{i, x} \leq e_j\}\]

Figure \ref{fig:overlap} visualizes the overlap region of a particular contribution when restricted to a single donor $u$. 

The following constraint ensures that we do not redistribute funds from a contribution to campaigns that have already ended or campaigns that the donor contributed to after the current campaign ended:

\[\forall i,j, \;\ R_{i, j} \leq \sum_{x \in \mathcal{O}_{i, j}} A_{i, x}\]

\begin{figure}[htb]
    \centering
    \includegraphics[width=1\linewidth]{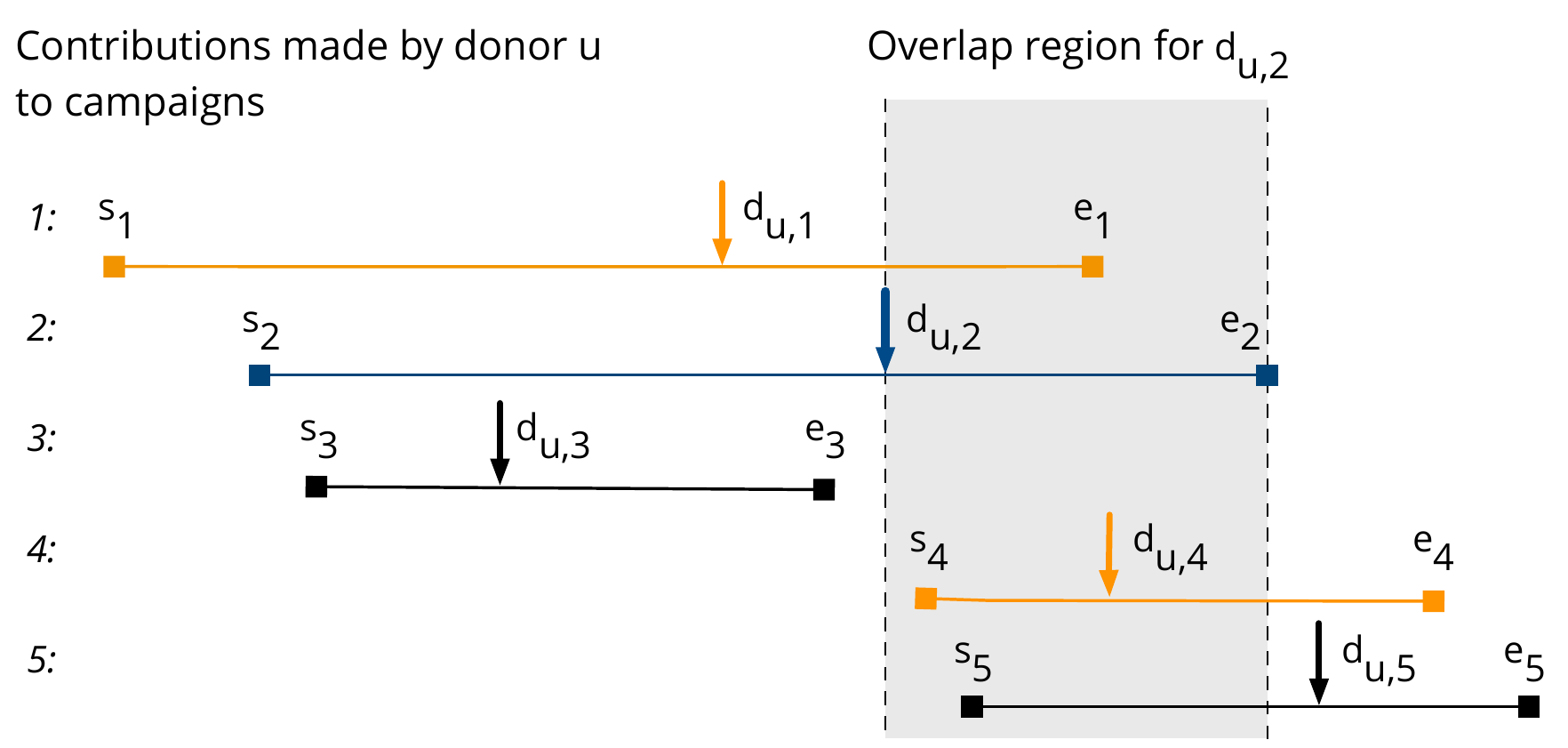}
    \caption{The contribution made by donor $u$ on $d_{u,2}$ for campaign $2$ overlaps with contributions made on $d_{u,1}$ and $d_{u,4}$. 
    For donor $u$, a redistribution of funds to campaign 2, can only be allowed from contributions made to campaign 1 and 4.}
    \label{fig:overlap}
\end{figure}

\end{enumerate}

\paragraph{Ordered \cpr}

Finally, the ordered \cpr scheme requires the following constraint:

\begin{enumerate}[label={[\bf\sc{Order}]},wide =1em]
\item \emph{Preserve Preference Ordering:} If a donor contributes more funds to one campaign compared to another, both absolutely and relatively to their respective goals, then a redistribution of funds should preserve this preference ordering.
\[\forall i,x,y, \;\  (A_{i, x} > A_{i, y}) \land (A_{i, x}/G_x > A_{i, y}/G_y) \\
\rightarrow  R_{i, x} \geq R_{i, y} \]

We relax preference ordering to eliminate contributions to \textit{failed} campaigns (i.e. campaigns that do not succeed for any feasible redistribution) with the following constraint instead:
\begin{multline*}
\forall i,x,y, \;\  (A_{i, x} > A_{i, y}) \land (A_{i, x}/G_x > A_{i, y}/G_y) \\
\land (I'_x = 1) \land (I'_y = 1) 
\rightarrow R_{i, x} \geq R_{i, y}
\end{multline*}
\end{enumerate}

%% file: discussion.tex
\section{Discussion}
\label{sec:discussion}

Even though our empirical analysis shows that redistribution leads to clear efficiency gains, successfully implementing redistribution within a crowdfunding platform is not without complications.  A na\"ive redistribution scheme can be both objectionable and detrimental. Taking funds from `rich' campaigns to give to `poor' ones may be seen as unfair and a violation of donor intentions.  Other secondary effects could also arise. The overall quality of funded campaigns may drop due to weaker projects being funded. If funders believe their funds may back projects that they do not deem worthy, they may also be less likely to contribute. In this section we discuss these and other challenges. We consider the incentives from the perspective of donors, campaign organizers, and platform designers and describe the implications of redistribution. 

\subsection{For Donors}

Donors' reasons and motivations for giving are diverse~\cite{Gerber:2013:CMD:2562181.2530540}. 
Donors are often emotionally invested in the campaigns they donate to: 
taking funds from one campaign and giving it to another may cause donors to feel slighted. 
Imagine donating \$50 to produce a film you are excited to see only to find that your funds were redistributed to make a potato salad that you cannot eat. Crowdfunding platforms could mitigate this reaction through the following approaches: (i) by asking backers up front whether they are willing to accept a redistribution, (ii) by respecting their giving preferences or (iii) by orienting a donor's mindset toward accepting total utilitarianism. Both offline giving clubs and DonorsChoose implement variants of our redistribution schemes and through a combination of the above approaches.

Within our categorization, offline giving clubs are similar in spirit to \cpr schemes. Giving clubs manage several campaigns that share a unified high-level cause such as education, environmental conservation, etc.. By re-framing the donation process as a collective action rather than an individualistic one, donors indicate their higher-level preference of the giving club's overall cause and either elected members or the majority dictates the distribution of the collected funds. 
On an online crowdfunding platform, similar campaigns can be grouped into giving themes with redistribution schemes automatically allocating funds across the grouped campaigns.

DonorsChoose implements a repurposing scheme, but is able to successfully do so through the combination of (i) having a cohesive theme of education, (ii) integrating redistribution into the charitable ethos of their platform, and (iii) vetting all projects. 

\subsection{For Campaign Organizers}

The question of reallocation on the part of organizers is similarly nuanced. The policy that additional funds given to a campaign may be reallocated to other (similar) campaigns may cause organizers to feel cheated of their raised funds. 

However, in the case of discrete public goods, the requested amount is determined by a specific need rather than of arbitrary size. In other words, campaign organizers should only be requesting the amount that they actually need; if they wanted to raise more, then they should ask for more initially since the amount requested for discrete public goods is based off of a cost estimate.

There are also very practical reasons why reallocation of excess funds can be \textit{good} for organizers. The most obvious benefit is if a campaign is unable to raise sufficient funds (over 50\% of Kickstarter's projects~\cite{mollick2014dynamics}), redistribution provides such a campaign with a better chance of success. Less obviously, Mollick et al. found evidence that projects with substantial (200\%) excess funds often deliver \textit{worse} results than those that just meet their goals~\cite{mollick2014dynamics}. Kickstarter's blog posted a series on how over-funded projects deal with the extra influx of funds~\cite{excess-blog}; organizers who find themselves in these situations are generally unprepared and must come up with strategies on the fly to cope with excess funds.

On the flip side, the existence of reallocation may also entice project organizers to try and game the system. The risk of not meeting goals and losing all funds is the same as before redistribution, but organizers now have an additional chance to meet their fundraising goal. Thus, organizers have an increased incentive to raise the fundraising goal hoping that reallocation provides an additional chance in their favor. These and other second-order effects and analysis of possible attacks against the system are beyond the scope of this paper. 

As with the donor's interface, the redistribution policy should be presented up front to the organizers so they can decide whether this is suited to their needs. DonorsChoose adopts this approach and further allows the organizers themselves to decide how to redistribute some of the funds they raised if they fail. This can give the organizers a sense of control, even if ultimately the moneys are given to other projects. This model appears to work well when all projects are vetted.

\subsection{For the Crowdfunding Designer}

As with any site-level changes, the crowdfunding platform designer must consider a wide-array of issues. These questions arise from donor and organizer perspectives, but also in relation to implementation on the site and meta-level implications for the platform's brand.
Modern crowdfunding campaigns align multiple incentives to motivate donors to contribute, including: progress updates, donor incentives (e.g. gifts and prizes), entertaining previews, and a sense of community. 
How then does redistribution interact with such incentives? One possibility is by completely automating the redistribution process the platform may lose some of the engagement that entices donors to return and give more. It is likely that there is no ``correct'' choice, but instead that donors fall along a spectrum of desired engagement levels.

Another question for crowdfunding designers is: when should the redistribution be conducted? Immediately after a campaign ends would require the consideration of other campaigns ending at similar times or across certain time windows. Alternatively, having large amounts of idle donations is wasteful and potentially problematic from the designer's standpoint.
There are also second-order effects to giving patterns that may emerge. For example, most donations occur at the start of a campaign and then towards the end of the campaign~\cite{wash-time} and it is satisfying for donors to see to a project's completion. Will redistribution dampen these effects or could we redistribute in a manner than enhances them?

Looking at the numbers from DonorsChoose, it is clear that they have successfully implemented the repurposing redistribution scheme by bringing together several elements that work well together. Once a crowdfunding platform successfully implements redistribution, the benefits could also increase as these policies become the norm. The effects of incentives, different ways of incorporating redistribution, and second-order effects are interesting questions worth examining in future work.

%% file: conclusion.tex
\section{Conclusion}

In this paper, we explore the possibility of redistribution of crowdfunding donations. We build the case for redistribution and develop a classification of redistribution models. We implement these models as a set of constraints within a campaign-success optimization problem and evaluate their potential efficiency using data from \lag, an online crowdfunding platform. We find that a range of redistribution schemes can provide different levels of benefits in terms of campaign success rates. We also find redistribution schemes in the wild and explore how they are able to succeed despite disregarding donor preferences. Finally, we considered different ways that redistribution mechanisms could be implemented and integrated within a crowdfunding platform and highlight some of the issues that crowdfunding designers should consider.

%% file: proceedings.bbl

\begin{thebibliography}{00}


\ifx \showCODEN    \undefined \def \showCODEN     #1{\unskip}     \fi
\ifx \showDOI      \undefined \def \showDOI       #1{{\tt DOI:}\penalty0{#1}\ }
  \fi
\ifx \showISBNx    \undefined \def \showISBNx     #1{\unskip}     \fi
\ifx \showISBNxiii \undefined \def \showISBNxiii  #1{\unskip}     \fi
\ifx \showISSN     \undefined \def \showISSN      #1{\unskip}     \fi
\ifx \showLCCN     \undefined \def \showLCCN      #1{\unskip}     \fi
\ifx \shownote     \undefined \def \shownote      #1{#1}          \fi
\ifx \showarticletitle \undefined \def \showarticletitle #1{#1}   \fi
\ifx \showURL      \undefined \def \showURL       #1{#1}          \fi

\bibitem{andreoni1988free}
{James Andreoni}. 1988.
\newblock \showarticletitle{Why free ride?: Strategies and learning in public
  goods experiments}.
\newblock {\em Journal of public Economics\/} {37}, 3 (1988), 291--304.
\newblock


\bibitem{belleflamme}
{Paul Belleflamme}, {Thomas Lambert}, {and} {Armin Schwienbacher}. 2014.
\newblock \showarticletitle{Crowdfunding: Tapping the right crowd}.
\newblock {\em Journal of Business Venturing\/} {29}, 5 (2014), 585--609.
\newblock


\bibitem{codo}
{Juan~Felipe Beltran}, {Aysha Siddique}, {Azza Abouzied}, {and} {Jay Chen}.
  2015.
\newblock \showarticletitle{Codo: Fundraising with Conditional Donations}. In
  {\em Proceedings of the 28th Annual ACM Symposium on User Interface Software
  Technology} {\em (UIST '15)}. ACM, New York, NY, USA, 213--222.
\newblock
\showISBNx{978-1-4503-3779-3}
\showDOI{%
\url{http://dx.doi.org/10.1145/2807442.2807509}}


\bibitem{cotterill2011impacts}
{Sarah Cotterill}, {Peter John}, {and} {Elizabeth Richardson}. 2011.
\newblock \showarticletitle{The impacts of a pledge campaign and the promise of
  publicity: A randomized controlled trial of charitable donations}.
\newblock {\em Available at SSRN 1833487\/} (2011).
\newblock


\bibitem{cumming2014crowdfunding}
{Douglas~J Cumming}, {Ga{\"e}l Leboeuf}, {and} {Armin Schwienbacher}. 2014.
\newblock \showarticletitle{Crowdfunding Models: Keep-it-All vs.
  All-or-Nothing}. In {\em Paris December 2014 Finance Meeting EUROFIDAI-AFFI
  Paper}.
\newblock


\bibitem{Gerber:2013:CMD:2562181.2530540}
{Elizabeth~M. Gerber} {and} {Julie Hui}. 2013.
\newblock \showarticletitle{Crowdfunding: Motivations and Deterrents for
  Participation}.
\newblock {\em ACM Trans. Comput.-Hum. Interact.\/} {20}, 6, Article 34 (Dec.
  2013), 32 pages.
\newblock
\showISSN{1073-0516}
\showDOI{%
\url{http://dx.doi.org/10.1145/2530540}}


\bibitem{greenberg2012crowdfunding}
{M Greenberg} {and} {E Gerber}. 2012.
\newblock \showarticletitle{Crowdfunding: A survey and taxonomy}.
\newblock  (2012).
\newblock


\bibitem{Hui:2014:ULS:2598510.2598539}
{Julie~S. Hui}, {Elizabeth~M. Gerber}, {and} {Darren Gergle}. 2014a.
\newblock \showarticletitle{Understanding and Leveraging Social Networks for
  Crowdfunding: Opportunities and Challenges}. In {\em Proceedings of the 2014
  Conference on Designing Interactive Systems} {\em (DIS '14)}. ACM, New York,
  NY, USA, 677--680.
\newblock
\showISBNx{978-1-4503-2902-6}
\showDOI{%
\url{http://dx.doi.org/10.1145/2598510.2598539}}


\bibitem{Hui:2014:URC:2531602.2531715}
{Julie~S. Hui}, {Michael~D. Greenberg}, {and} {Elizabeth~M. Gerber}. 2014b.
\newblock \showarticletitle{Understanding the Role of Community in Crowdfunding
  Work}. In {\em Proceedings of the 17th ACM Conference on Computer Supported
  Cooperative Work \&\#38; Social Computing} {\em (CSCW '14)}. ACM, New York,
  NY, USA, 62--74.
\newblock
\showISBNx{978-1-4503-2540-0}
\showDOI{%
\url{http://dx.doi.org/10.1145/2531602.2531715}}


\bibitem{kickstarterstats}
{{Kickstarter PBC}}. 2016.
\newblock Stats.
\newblock   (2016).
\newblock
\showURL{%
\url{https://www.kickstarter.com/help/stats}}


\bibitem{complete-effect}
{David Klinowski}, {Nichole Argo}, {and} {Tamar Krishnamurti}. 2015.
\newblock The completion effect in charitable crowdfunding.
\newblock   (2015).
\newblock
\showURL{%
\url{http://pitt.edu/~djk59/klinowski_argo_krishnamurti_crowdfunding.pdf}}


\bibitem{Mitra:2014:LGP:2531602.2531656}
{Tanushree Mitra} {and} {Eric Gilbert}. 2014.
\newblock \showarticletitle{The Language That Gets People to Give: Phrases That
  Predict Success on Kickstarter}. In {\em Proceedings of the 17th ACM
  Conference on Computer Supported Cooperative Work} {\em (CSCW '14)}. ACM, New
  York, NY, USA, 49--61.
\newblock
\showISBNx{978-1-4503-2540-0}
\showDOI{%
\url{http://dx.doi.org/10.1145/2531602.2531656}}


\bibitem{mollick2014dynamics}
{Ethan Mollick}. 2014.
\newblock \showarticletitle{The dynamics of crowdfunding: An exploratory
  study}.
\newblock {\em Journal of Business Venturing\/} {29}, 1 (2014), 1 -- 16.
\newblock
\showISSN{0883-9026}
\showDOI{%
\url{http://dx.doi.org/10.1016/j.jbusvent.2013.06.005}}


\bibitem{excess-blog}
{Niina Pollari}. 2015.
\newblock So Your Project Blew Up. Now What?
\newblock   (2015).
\newblock
\showURL{%
\url{https://www.kickstarter.com/blog/so-your-project-blew-up-now-what-part-one}}


\bibitem{wash-time}
{Jacob Solomon}, {Wenjuan Ma}, {and} {Rick Wash}. 2015.
\newblock \showarticletitle{Don't Wait!: How Timing Affects Coordination of
  Crowdfunding Donations}. In {\em Proceedings of the 18th ACM Conference on
  Computer Supported Cooperative Work \&\#38; Social Computing} {\em (CSCW
  '15)}. ACM, New York, NY, USA, 547--556.
\newblock
\showISBNx{978-1-4503-2922-4}
\showDOI{%
\url{http://dx.doi.org/10.1145/2675133.2675296}}


\bibitem{wash-skew}
{Jacob Solomon}, {Wenjuan Ma}, {and} {Rick Wash}. 2016.
\newblock \showarticletitle{Highly Successful Projects Inhibit Coordination on
  Crowdfunding Sites}. In {\em Proceedings of the 2016 CHI Conference on Human
  Factors in Computing Systems} {\em (CHI '16)}. ACM, New York, NY, USA,
  4568--4572.
\newblock
\showISBNx{978-1-4503-3362-7}
\showDOI{%
\url{http://dx.doi.org/10.1145/2858036.2858163}}


\bibitem{wash-returnrule}
{Rick Wash} {and} {Jacob Solomon}. 2014.
\newblock \showarticletitle{Coordinating Donors on Crowdfunding Websites}. In
  {\em Proceedings of the 17th ACM Conference on Computer Supported Cooperative
  Work \&\#38; Social Computing} {\em (CSCW '14)}. ACM, New York, NY, USA,
  38--48.
\newblock
\showISBNx{978-1-4503-2540-0}
\showDOI{%
\url{http://dx.doi.org/10.1145/2531602.2531678}}


\bibitem{xu2014show}
{Anbang Xu}, {Xiao Yang}, {Huaming Rao}, {Wai-Tat Fu}, {Shih-Wen Huang}, {and}
  {Brian~P Bailey}. 2014.
\newblock \showarticletitle{Show me the money!: An analysis of project updates
  during crowdfunding campaigns}. In {\em Proceedings of the SIGCHI Conference
  on Human Factors in Computing Systems}. ACM, 591--600.
\newblock


\end{thebibliography}
